\documentclass[aps,prl,preprint,groupedaddress,draft,showpacs, amssymb,amsmath]{revtex4-1}
\newcommand{\tr}{\mathop{\textrm {Tr}}\nolimits}
\begin{document}


\title{Proof of Bose condensation for weakly interacting lattice bosons}


\author{D. P. Sankovich}
\email{sankovch@mi.ras.ru}
\affiliation{V. A. Steklov Mathematical Institute, Gubkin str. 8, Moscow, Russia.}
\date{\today}
\begin{abstract}
A weakly interacting Bose gas on a simple cubic lattice is considered.
We prove the existence of the standard or zero-mode Bose condensation at sufficiently low temperature.
This result is valid for sufficiently small interaction potential and small values of chemical potential.
Our method exploits infrared bound for the suitable two-point Bogolyubov's inner product.
We do not use the reflection positivity or some expansion methods.
\end{abstract}


\maketitle

\section{\label{sec:Introduction}Introduction}

Bose condensation (BC) is one of the most amazing phenomena exhibited by macroscopic systems.
The study of condensation is fundamental because it is a clue to our understanding of phase transitions.
BC was first described for an ideal gas of free bosons.
In three spatial dimensions, BC occurs at low enough temperature when there is a nonzero density of particles in zero-momentum state:
\begin{equation}
\label{0}
n_0=\lim_{V\to\infty}\frac1V\left\langle b_0^\dag b_0\right\rangle>0,
\end{equation}
where
$b_k^\dag,b_k$
are the creation and annihilation operators for the single-particle state of momentum 
$\hbar k$,
$V$ 
is the volume of system and
$\left\langle\dots\right\rangle$
denotes thermal averaging.
It is recognized that the BC is a common quantum property of the many-particle systems.
For interacting systems the standard criterion (\ref{0}) was reexamined.
Onsager and Penrose \cite{OP} proposed to identity condensation with an off-diagonal long-range order in the one-particle reduced density matrix.
This criterion shows that the thermal average of particle operator for the mode $p=0$ can still be used as a characterization of a BC (in this case the thermal average in (\ref{0}) is taken over interacting Hamiltonian.)
Subsequently, a more detailed classification of the different types of BC was proposed \cite{BLP}.
We will rely on the standard definition.
According to the classification of the work \cite{BLP}, this type of BC is called the conventional condensation of type I.

The experimental creation of BC (for a review of the theory of trapped Bose gases with extensive references in the literature see \cite{AS1998}) has sparked interest in their properties.
One of the most hard and comprehensive problem is connected with the rigorous proof of the existence (or absence) of BC for realistic non-ideal Bose systems.
However it is believed that this the solvable issue.
We will not dwell on the many aspects of BC and refer the reader to monographs \cite{CPethick2008,LPitaevskii2008,EHLieb2003,AGriffin2009,AGriffin1995}.
In the following we consider a gas of interacting bosons in a lattice.
Let us consider this case in more detail.

Lattice systems in theoretical physics has long been regarded as an idealization of a natural crystal, subsystems which can be in a finite number of states \cite{DRuelle1969}.
Mathematical physics applies lattice models to approximate the quantum fields and make sense of the various formal considerations \cite{BSimon1974}.
After 1995, the development of experimental physics \cite{MHAnderson1995,CCBradley,KBDavis1995} allowed to look at the system as a grid of the observed physical objects, allowing for a convenient practical implementation \cite{AS1998}.
The successes in the experimental study of BC are numerous and impressive.
Nevertheless, the rigorous theoretical justification for this phenomenon remains, as before, not completely solved, even for weakly interacting systems with pair interaction.

The basic model of interacting lattice bosons is the Bose--Hubbard (BH) model \cite{HA1963,MPAFisher1989}.
The possibility of applying this model to gases of alkali atoms in optical traps was first realized in \cite{DJaksch1998}.
Theoretical investigation of phase transitions in the BH model is mainly based on the application of some or other approximate or numerical methods.
We will not discuss these methods and consider a small number of rigorous results.
Note that all these results relate mainly to some simplifying modifications of the BH model (see a very complete review \cite{U2002}.)
In \cite{JB2003} the lattice infinite-range hopping BH model has been studied for all temperatures and chemical potentials.
A model with a hard-core BH potential was studied by rigorous perturbation theory in \cite{RFernandez2006}.
A related model with extra chessboard potential is considered in \cite{MAizenman2004}.
The authors of the last article used the equivalence of the model considered by them and the 
$XY$
model of spin 
$1/2$
in a magnetic field.
This model made known property of Gaussian domination \cite{JFrohlich1976,FJDyson1976,JFrohlich} that allows us to prove the presence of BC.
Recently the upper bound on the isothermal compressibility for lattice bosons in the uniform BH model was derived \cite{DP}.
Consideration in this article does not exploit reflection positivity and infrared bounds or some expansion techniques.

In this paper we investigate BC in the uniform BH model.
It represents a simple lattice model of wandering bosons which interact locally.
We focus on the situation with sufficiently small positive interaction parameter.
In other words, the case of a weakly interacting Bose gas on a lattice is considered.
As noted above, the existence of a BC was proved for the hard-core Bose gas where each site can be occupied by at most one particle (infinite interaction parameter) \cite{MAizenman2004,RFernandez2006}.
The method we use is based on the condition of Gaussian domination.
However, unlike previously known works, we do not use the property of reflection positivity to obtain the necessary infrared bound.
The possibility of applying infrared estimates for a rigorous proof of the BC was advanced in 1980 \cite{JFrohlich1980}.
For various model Bose systems, the method of infrared bounds was used earlier in \cite{DP1989,MCorgini2002}.

The paper is organized as follows.
In Section 2, we introduce the model and obtain the necessary upper bound for the suitable Bogolyubov's inner product (Duhamel two-point function) \cite{RKubo1957,NN1962,GRoepstorff1976}.
The proof of the existence of BC is presented in Section 3.
The conditions under which this condensation is possible are also given there.
Finally, we end with a conclusion and outlook in Section 4.

\section{\label{sec:model} Model and gaussian domination}

We consider a many boson system in equilibrium, at a given
temperature 
$T$ ($\beta=(kT)^{-1},$ where $k$ is the Boltzmann constant), a given chemical potential $\mu$, and with
given interactions.
The system is described in the grand canonical formalism.  
Let us be more precise and introduce the mathematical framework.
Let $\Lambda\subset\mathbb{Z}^3$ be a finite cube of volume $V=|\Lambda|$. 
Introduce the bosonic Fock space $\mathcal{F}=\oplus_{N\geq0}\mathcal{H}_{\Lambda,N}$,
where $\mathcal{H}_{\Lambda,N}$
is the Hilbert space of symmetric complex functions on $\Lambda^N$. 
Creation and annihilation operators for a boson at site $j\in\Lambda$ 
are denoted by
$a^{\dag}_j$ and $a_j$, respectively.
Then $n_j=a^{\dag}_j a_j$  is the one-site number operator, and
$N_\Lambda=\sum_{j\in\Lambda}n_j$
is the total number operator.

The basic Hamiltonian of the uniform Bose--Hubbard model is
\begin{align}
\label{1}
H_\Lambda =&\frac{t}{2}\sum_{n,i}(a^\dag_n-a^\dag_{n+\delta_i})(a_n-a_{n+\delta_i})\notag\\
&+\frac U2\sum_n a^\dag_n a_n(a^\dag_n a_n-1)-\mu N_\Lambda,
\end{align}
where $i$ is summed from $1$ to $3$, and $n$ is summed over $\Lambda$.
Here $\delta_i$ is the unit vector whose $i$-th component is $1$.
We shall consider the periodic boundary conditions, so
$$
  \Lambda=\{n\in\mathbb{Z}^3:-L_i/2\leq n_i<L_i/2, i=1,2,3\}
$$
is a domain of $\mathbb{Z}^3$ wrapped onto a torus.
Then the set
$$
	\Lambda^*=\{k_i=2\pi l/{L_i}:l=0,\pm1,\ldots,\pm L_i/2, i=1,2,3\}
$$
is dual to $\Lambda$ with respect to Fourier transformation on the domain
$
\Lambda=L_1\times L_2\times L_3$.
The first term in the Hamiltonian (\ref{1}) corresponds to the hopping interaction of bosons between neighboring sites. 
The hopping parameter $t$ is chosen to be positive.
The second term in (\ref{1}) is the on-site repulsive interaction ($U>0$).
The Hamiltonian (\ref{1}) is superstable. 
Indeed, it is not hard to see \cite{DP} that
$$
H_\Lambda\geq\frac{U}{2|\Lambda|}N^2_\Lambda-\left(\mu+\frac{U}{2}\right)N_\Lambda.
$$
The superstability condition implies the convergence of the grand-partition function for any 
$
\mu\in\mathbb {R}^1,\beta>0
$.
Notice that attraction ($U<0$) makes the model unstable (in contrast to the fermion case.)

Let $b^\dag_p,b_p$ are creation and annihilation Bose operators with the wave vector $p\in\Lambda^*$,
$$
b^\dag_p=\frac1{\sqrt{|\Lambda|}}\sum_{n\in\Lambda}a^\dag_n e^{i pn}, b_p=\frac1{\sqrt{|\Lambda|}}\sum_{n\in\Lambda}a_n e^{-ipn}.
$$
In terms of $b^\dag_p,b_p$, the Hamiltonian $H_\Lambda$ becomes
$$
H_\Lambda= \sum_p \omega_p b^\dag_p b_p+ \frac{U}{2|\Lambda|}\sum_{p,q,k}b^\dag_p b^\dag_q b_{p+k}b_{q-k}-\mu N_\Lambda,
$$
where
$
N_\Lambda=\sum_p b^\dag_p b_p
$
and 
$$
\omega_p= 4t\sum_{\alpha=1}^3 \sin^2\frac {p_\alpha}2\equiv t\epsilon_p\geq 0.
$$
Since the summation of $p,q,k$ is always restricted to $\Lambda^*$, we will not explicitly specify it.

First, we define the Bogolyubov inner product 
$$
\left(b^\dag_p, b_p\right)=Z_\Lambda^{-1}\int_0^1 {\tr}\left(b^\dag_p e^{-x\beta H_\Lambda} b_p e^{-(1-x)\beta H_\Lambda}\right)dx,
$$
where
$
Z_\Lambda={\tr }\exp(-\beta H_\Lambda)
$
is the grand-canonical partition function.
Note, that the thermal expectation of the double commutator
$$
c_p\equiv \left\langle[b^\dag_p,[\beta H_\Lambda,b_p]]\right\rangle_{H_\Lambda}\geq0.
$$
This follows from Bogolyubov's inequality or by an eigenfunction expansion \cite{DLS}.
The non-negativity of $c_p$ means
\begin{equation}
\label{2}
\mu\leq 2Un_\Lambda,
\end{equation}
where 
$
n_\Lambda=\left\langle N_\Lambda\right\rangle_{H_\Lambda}/|\Lambda|
$.

Consider the family of operators
\begin{align*}
\label{3}
H_\Lambda(h)=&\sum_p \omega_p(b^\dag_p-h^*_p)(b_p-h_p)\notag\\
&+
\frac{U}{2|\Lambda|}\sum_{p,q,k}b^\dag_p b^\dag_q b_{p+k}b_{q-k}-\mu N_\Lambda
\end{align*}
and introduce the function
\begin{equation}
\label{4}
f(U)=\tr e^{-\beta H_\Lambda(h)}-\tr e^{-\beta H_\Lambda(0)},
\end{equation}
where $h_p\in\mathbb{C}$.
We will explore the region of sufficiently small $U$.
In virtue of inequality (\ref{2}) let us focus on the situation with a small and non-negative chemical potential, 
$\mu=U\lambda+o(U)$, where $\lambda\geq0$.
It is easy to verify that $f(0)=0$. 
The derivative of $f(U)$ with respect to $U$ is
\begin{widetext}
\begin{equation}
\label{5}
f'(U)=-\beta\tr\left[\left(\frac{1}{2|\Lambda|}\sum_{p,q,k}b^\dag_p b^\dag_q b_{p+k}b_{q-k}-\lambda\sum_p b^\dag_p b_p\right)\left(e^{-\beta H_\Lambda(h)}
-e^{-\beta H_\Lambda(0)}\right)\right].
\end{equation}
\end{widetext}
By (\ref{5}), we obtain
\begin{widetext}
\begin{equation}
\label{6}
\noindent
f'(0)= -\beta\tr\left(\frac{1}{2|\Lambda|}\sum_{p,q,k}h^*_p h^*_q h_{p+k}h_{q-k}+
\frac{2}{|\Lambda|}\sum_p h^*_ph_p N_\Lambda-\lambda\sum_p h^*_ph_p\right)e^{-\beta H_\Lambda^{(0)}},\nonumber
\end{equation}
\end{widetext}
where
$$
H_\Lambda^{(0)}=\sum_p \omega_p b^\dag_p b_p.
$$
This implies that $f'(0)\leq0$ for $\lambda\leq2n^{(0)}$, where $n^{(0)}=\left\langle N_\Lambda\right\rangle_{H_\Lambda^{(0)}}/|\Lambda|$.
We conclude that $f(U)\leq0$ for $U$ and $\mu$ small enough.
The inequality $f(U)\leq0$ is the Gaussian domination \cite{JFrohlich1976}.

\section{\label{sec:BC}Bose condensation}

In this section we want to put the results of the previous section to prove that the BC occurs in the model (\ref{1}) for $U$ and $\mu$ sufficiently small.

From the inequality 
$
f(U)\leq0
$
, one concludes that 
$
\tr\exp(-\beta H_\Lambda(h))
$ 
takes the maximum value at $\{h_p=0\}$.
A necessary condition for 
$
\tr\exp(-\beta H_\Lambda(h))
$ 
to be maximum at $\{h_p=0\}$ is represented by inequality 
\begin{equation}
\label{7}
\left(b^\dag_p,b_p\right)_{H_\Lambda}\leq{(\beta \omega_p)}^{-1},\; p\neq0.
\end{equation}

Infrared bound (\ref{7}) is essential for our proof of the BC.
The proof comes from two points.
The first is the Falk--Bruch inequality \cite{FB}.
We use this inequality to relate Bogolyubov's inner product to the conventional two-point thermal average.
The second is the sum rule for this average.
Using this rule, we obtain some conditions, in which the contribution from $p=0$ remains non-vanishing in the thermodynamic limit.
As a result, we prove the existence of one-mode Bose condensate.
This is the method of infrared bounds that was originally introduced for the classical Heisenberg model in \cite{JFrohlich1976}.
Later on this method was extended to the quantum spin systems \cite{DLS} and to the Bose systems \cite{DS1,DP1989}.

As noted above, the upper bound for the two-point temperature average
$\left\langle b^\dag_p b_p\right\rangle_{H_\Lambda}$
can be obtained from the upper bound of 
$(b^\dag_p, b_p)_{H_\Lambda}$
by the Fulk--Bruch inequality.
For any $A$ and self-adjoint $H$ we have the bound 
$$
b(A)\geq g(A)h\left(\frac{c(A)}{4g(A)}\right),
$$
where 
\begin{align*}
b(A)&\equiv(A^\dag,A),g(A)\equiv\frac12\langle A^\dag A+AA^\dag\rangle,\\
c(A)&\equiv\langle[A^\dag,[\beta H,A]]\rangle,
\end{align*}
and 
$
h(x\tanh x)=x^{-1}\tanh x.
$
The function $h$ is a well defined strictly monotonically decreasing convex function from $(0,\infty)$ to $(0,1)$ with
$$
\lim_{x\to 0} h(x)=1,\,\,\lim_{x\to\infty} h(x)=0.
$$
Suppose that $b\leq b_0$ and $c\leq c_0$.
Then $g\leq g_0$, where
$$
g_0=\frac12\sqrt{c_0b_0}\coth\sqrt{\frac{c_0}{4b_0}}.
$$
We refer to \cite{DLS} for the proof of these relations, and for further statements about correlation functions.
The thermal average of the double commutator is
\begin{equation}
\label{9}
c(b_p)\equiv c_{p}=\beta(\omega_{{p}}+2n_\Lambda U-\mu),\,\, {p}\in\Lambda^*,
\end{equation}
where 
$n_\Lambda$ is the filling (the thermal average number of particles per site.)
The non-negativity of $c_{p}$ means
$$
\Delta_\Lambda\equiv2n_{\Lambda}U-\mu\geq0.
$$
From the Falk--Bruch inequality and bounds (\ref{7}), (\ref{9}) we infer that for ${p}\neq0$,
\begin{widetext}
\begin{equation}
\label{11}
\left\langle b^{\dag}_{p}b_{p}\right\rangle_{H_\Lambda}\leq\frac12\sqrt{\frac{\omega_{p}+2n_\Lambda U-\mu}{\omega_{p}}}
\coth\frac{\beta}{2}\sqrt{\omega_{p}(\omega_{p}+2n_\Lambda U-\mu)}-\frac{1}{2}\equiv F_\Lambda(p,\mu).\nonumber
\end{equation}
\end{widetext}
From the sum rule 
\begin{equation}
\label{sr}
\frac1{|\lambda|}\sum_{{p}}\left\langle b^{\dag}_{p}b_{p}\right\rangle_{H_\Lambda}=\frac{\langle N\rangle_{H_\Lambda}}{|\Lambda|}=n_\Lambda,
\end{equation}
we conclude that the Bose condensate density
$$\label{13}
n_0=\lim_{|\Lambda|\rightarrow\infty}\frac1{|\Lambda|}\langle b^{\dag}_{0}b_{0}\rangle_{H_\Lambda}
$$
remains non-vanishing in the thermodynamical limit if
\begin{equation}
\label{14}
n=\lim_{|\Lambda|\rightarrow\infty}n_\Lambda>\frac{1}{16\pi^3}\int_{-\pi}^\pi dk_1\int_{-\pi}^\pi dk_2\int_{-\pi}^\pi dk_3\,\, F(k,\mu),
\end{equation}
where
$F({p},\mu)=\lim_{|\Lambda|\rightarrow\infty}F_\Lambda({p},\mu)$.

We will explore the region of non-negative chemical potentials.
In this case we have
$
\Delta=\lim_{|\Lambda|\rightarrow\infty}\Delta_\Lambda\leq2nU.
$
Inequality (\ref{14}) is then more accomplished if will do the following inequality
\begin{equation}
\label{15}
n>\frac{1}{16\pi^3}\int_{-\pi}^\pi dk_1\int_{-\pi}^\pi dk_2\int_{-\pi}^\pi dk_3\,\, F({k},0).
\end{equation}
This follows from the fact that the function $F({p},\mu)$ is a monotonically increasing function of $\Delta$.

Derive the conditions under which executes the inequality (\ref{15}).
Consider first the case of zero temperature.
Then
\begin{widetext}
$$
n>\frac1{16\pi^3}\int_{-\pi}^\pi dk_1\int_{-\pi}^\pi dk_2\int_{-\pi}^\pi dk_3\,\,\left(\sqrt{1+\frac{2nU}{\omega_{k}}}-1\right)\equiv J_0(n).
$$
\end{widetext}
If $J_0^{'}(0)<1$, then $J_0(n)<n$ for all $n\geq 0$.
Inequality  $J_0^{'}(0)<1$ is performed if
\begin{equation}
\label{16}
\frac{U}t\leq\frac{12}{W_0},
\end{equation}
where $W_0\approx 1.51$ is the Watson's integral.

Consider the case of positive temperatures.
Use the estimate $\coth x\leq1+x^{-1}$ in the main inequality (\ref{15}).
Then we have that (\ref{15}) is accomplished if
\begin{equation}
\label{17}
n-J_0(n)>\frac{W_0}{3\beta t}\,.
\end{equation}
We conclude that there is Bose condensation at some finite $\beta$ whenever (\ref{16}) and (\ref{17}) hold.
The temperature of the phase transition is estimated from below as 
$$\label{18}
\theta_c\geq\frac{3t}{W_0}[n-J_0(n)].
$$ 
\section{\label{sec:conc}Conclusion}

In this paper we have studied a lattice superstable model of imperfect Bose gas.
The presence of a BC is established for small enough interaction potential $U$, and small chemical potential $\mu\geq0$.
A lower estimate for the critical temperature is received.
Our proofs exploit infrared bounds, and does not exploit reflection positivity or some expansion methods.
The essential ingredient in the proof of main results is the basic bound (\ref{7}).
Our method relies on the fact that the bound (\ref{7}) and the sum rule (\ref{sr}) force a macroscopic occupation in the $p=0$ mode.
This is the conventional BC.

We proved (\ref{7}) for small enough $U$.
Note that (\ref{7}) is true for small enough $t$.
Indeed, consider (\ref{4}) as function of $t$.
Then
$$
f'(t=0)=-\beta\tr\left[\sum_p\epsilon_p|h_p|^2 e^{-\beta\left(H_\Lambda|_{t=0}-\mu N_\Lambda\right)}\right]\leq0,
$$
for any $\mu$. 
We see that (\ref{7}) is true for $t$ small enough ($U$ is large enough.)
Will (\ref{7}) be true for any $U$ and $t$?
This fundamental issue remains open.
\bibliography{proofBC}

\end{document}